Factor endowment–commodity output relationships in a three-factor two-good general equilibrium trade model: Further analysis[1]

Written on Oct. 1, 2018

Authors: Yoshiaki Nakada

Division of Natural Resource Economics, Graduate school of Agriculture, Kyoto University, Sakyo, Kyoto, Japan. E-mail: nakada@kais.kyoto-u.ac.jp

Abstract:

The position of the EWS (economy-wide substitution)-ratio vector determines the Rybczynski sign pattern, which expresses the factor endowment–commodity output relationships in a three-factor two-good general equilibrium trade model. In this article, we show that the EWS-ratio vector exists on the line segment. Using this relationship, we develop a method to estimate the position of the EWS-ratio vector. We derive a sufficient condition for extreme factors to be economy-wide complements, which implies 'a strong Rybczynski result.' Additionally, we derive a sufficient condition for a specific Rybczynski sign pattern to hold. We assume factor-intensity ranking is constant.

JEL D50, D58, F10, F11

Introduction

Batra and Casas (1976) (hereinafter BC) wrote an article on functional relations in a three-factor two-good neoclassical model (hereinafter 3 x 2 model) and claimed that 'a strong Rybczynski result' arises. However, this is not the case (see Nakada (2017)). According to Suzuki (1983, p. 141), BC contended in Theorem 6 (p. 34) that 'if commodity 1 is relatively capital intensive and commodity 2 is relatively labor intensive, an increase in the supply of labor increases the output of commodity 2 and reduces the output of commodity 1. [Moreover, an increase in the supply of capital increases the output of commodity 1 and reduces the output of commodity 2.]' This is what a strong Rybczynski result implies.

---

[1] An earlier version was titled, "Factor endowment–commodity output relationships in a three-factor two-good general equilibrium trade model: Further analysis," uploaded by Yoshiaki Nakada on 31 May 2016. Available at
https://www.researchgate.net/publication/303683776_Factor_endowment-commodity_output_relationships_in_a_three-factor_two-good_general_equilibrium_trade_model_Further_analysis

Nakada (2017) defined the EWS-ratio vector based on 'economy-wide substitution' (hereinafter EWS) originally defined by Jones and Easton (1983) (hereinafter JE) and used it in an analysis. Nakada (2017) concluded that the position of the EWS-ratio vector determines the Rybczynski sign pattern, which expresses the factor endowment–commodity output relationships and its dual counterpart, the Stolper-Samuelson sign pattern, which expresses the commodity price–factor price relationships in a 3 x 2 model of BC's original type. Using the EWS-ratio vector, Nakada (2017) derived a sufficient condition for 'a strong Rybczynski result' to hold (or not to hold). The following result has been established (see Theorem A.1).

Theorem A.1 …Further, if the EWS-ratio vector $(S', U')$ exists in quadrant IV (or subregions P1–P3), in other words, if extreme factors are economy-wide complements, a strong Rybczynski result necessarily holds….

For subregions P1–P3, see Fig. A1 in Appendix A. The following questions arise.
(i) How can we estimate the position of the EWS-ratio vector?
(ii) Under what conditions does the EWS-ratio vector exist in quadrant IV; in other words, are extreme factors economy-wide complements?

The purpose of this paper is as follows. First, we show that the EWS-ratio vector exists on the line segment. Using this relationship, we develop a method to estimate the position of the EWS-ratio vector. We derive a sufficient condition for the EWS-ratio vector exist in quadrant IV. If this holds, three of the Rybczynski sign patterns hold (see Theorem A.1). Additionally, we derive a sufficient condition for the EWS-ratio vector to exist in a specific subregion. If this holds, a specific Rybczynski sign pattern holds. This article will provide the basis for further applications.

Similar to Nakada (2017), some papers are interested in the role of complementarity. For example, Takayama (1982, p. 19) assumed that extreme factors were aggregate complements. Suzuki (1987, Chapter 1, p. 17–26) assumed that extreme factors were Allen complements in each sector. Teramachi (1993) assumed that extreme factors were aggregate complements. On the other hand, Thompson (1985) assumed that any two factors were aggregate complements.[2] Other papers also address complementarity, for example, Thompson (1995), Bliss (2003), Easton (2008), Ide (2009), Ban (2007a), and Ban (2008). See also Ban (2007b) and Ban (2011, chapter 4, p. 87-9).[3] For details, see the Introduction in Nakada (2017).

---

[2] In the Appendix (p. 66-70), Teramachi (1993) commented that the analysis in Thompson (1985) was implausible.
[3] Additionally, Suzuki (1983) assumed that capital and land (middle factor and extreme factor, respectively) were 'perfect complements' in each sector, and derived the implications using Allen-partial elasticities of substitution. JE assumed that two factors were 'perfect complements' and derived the implications using EWS (see JE (p. 90–92)). However, Suzuki's proof is implausible (see Nakada

In summary, some of these previous studies are somewhat complicated. We are uncertain whether all of these studies are plausible. At least, to the author's knowledge, none of the studies analyze the conditions under which extreme factors are economy-wide complements (or aggregate complements, Allen complements).

Section 2 of this study explains the model. In subsection 2.1, we explain the basic structure of the model. In subsection 2.2, we assume factor-intensity ranking. In subsection 2.3, we derive the important relationship between EWS-ratios and draw the EWS-ratio vector boundary. In section 3, we estimate the position of the EWS-ratio vector under some assumptions. In subsection 3.1, we show that the EWS-ratio vector is on the line segment *AB* (or the EWS-ratio vector line segment). In subsection 3.2, we define the factor-price-change ranking and its implications. In subsection 3.3, by analyzing the Cartesian coordinates of points *A* and *B*, we develop a method to estimate the position of the EWS-ratio vector. First, we derive a sufficient condition for the EWS-ratio vector to exist in quadrant IV, that is, in any subregion P1, P2, or P3. Moreover, we derive a sufficient condition for the EWS-ratio vector to exist in a specific subregion. Finally, I show some implications. Section 4 presents the conclusions. In Appendix A, I show the details of analysis in Nakada (2017). In Appendix B, we show that the EWS-ratio vector exists on the straight line, called the EWS-ratio vector line. In Appendix C, we prove Lemma 1. In Appendix D, we derive the important relationships among some variables. Next, using these relationships, we prove Lemma 2. In Appendix E, we prove Theorem 1. In Appendix F, we investigate the assumption that production functions are of the two-level CES type. Section 2 contains similar content as Nakada (2017).

2. Model
 2.1. Basic structure of the model

We assume similarly to BC (p.22-23). That is, we assume as follows. Products and factors markets are perfectly competitive. Supply of all factors is perfectly inelastic. Production functions are homogeneous of degree one and strictly quasi-concave. All factors are not specific and perfectly mobile between sectors, and factor prices are perfectly flexible. These two ensure the full employment of all resources. The country is small and faces exogenously given world prices, or the movement in relative price of a commodity is exogenously determined. The movements in factor endowments are exogenously determined.

Full employment of factors implies

$$\sum_j a_{ij} X_j = V_i, \ i = T, K, L,  \tag{1}$$

---

(2015)). JE's proof in subsections 5.2.4, and 5.2.5 (p. 90–92) is questionable (see Nakada (2017, Appendix)).

where $X_j$ denotes the amount produced of good $j$ ($j$=l, 2); $a_{ij}$ the requirement of input $i$ per unit of output of good $j$ (or the input-output coefficient); $V_i$ the supply of factor $i$; $T$ is the land, $K$ capital, and $L$ labor.

In a perfectly competitive economy, the unit cost of production of each good must just equal to its price. Hence,

$$\sum_i a_{ij} w_i = p_j, \ j = 1, 2, \qquad (2)$$

where $p_j$ is the price of good $j$; $w_i$ is the reward of factor $i$.

BC (p.23) stated, 'With quasi-concave and linearly homogeneous production functions, each input-output coefficient is independent of the scale of output and is a function solely of input prices:'

$$a_{ij} = a_{ij}(w_i), \ i = T, K, L, \ j = 1, 2. \qquad (3)$$

And they continued, 'In particular, each $C_{ij}$ [$a_{ij}$ in our expression] is homogeneous of degree zero in all input prices.'[4]

Equations (1)-(3) describe the production side of the model. These are equivalent to eqs (1)-(5) in BC. The set includes 11 equations in 11 endogenous variables ($X_j$, $a_{ij}$, and $w_i$) and five exogenous variables ($V_i$ and $p_j$). The small-country assumption simplifies the demand side of the economy. Totally differentiating eq. (1), we have

$$\Sigma_j (\lambda_{ij} a_{ij}^* + \lambda_{ij} X_j^*) = V_i^*, \ i = T, K, L, \qquad (4)$$

where an asterisk denotes a rate of change (e.g., $X_j^* = dX_j/X_j$), and where $\lambda_{ij}$ is the proportion of the total supply of factor $i$ in sector $j$ (that is, $\lambda_{ij} = a_{ij} X_j / V_i$). Note that $\Sigma_j \lambda_{ij} = 1$.

The minimum-unit-cost equilibrium condition in each sector implies $\Sigma_i w_i da_{ij} = 0$. Hence, we derive (see JE (p.73, eq. (9)), BC (p.24, note 5),

$$\Sigma_i \theta_{ij} a_{ij}^* = 0, \ j = 1, 2, \qquad (5)$$

where $\theta_{ij}$ is the distributive share of factor $i$ in sector $j$ (that is, $\theta_{ij} = a_{ij} w_i / p_j$.). Note that $\Sigma_i \theta_{ij} = 1$; $da_{ij}$ is the differential of $a_{ij}$.

---

[4] From the condition of cost minimization, we can show that $a_{ij}$ is homogeneous of degree zero in all input prices (see Samuelson (1953, chapter 4, p. 68), Nakada (2017, p. 5)).

Totally differentiating eq. (2), we derive

$$\Sigma_i \theta_{ij} w_i^* = p_j^*, \; j=1, \; 2. \tag{6}$$

Subtracting $p_1^*$ from the both sides of eq. (6), we have

$$\Sigma_i \theta_{i1} w_{i1}^* = 0,$$
$$\Sigma_i \theta_{i2} w_{i1}^* = -P, \tag{7}$$

where $P = p_1^* - p_2, w_{i1}^* = w_i^* - p_1^* = (w_i / p_1)^*$; $P$ is the rate of change in the relative price of a commodity; $w_{i1}$ is the real factor price measured by the price of good 1.

Totally differentiate eq. (3) to obtain

$$a_{ij}^* = \Sigma_h \varepsilon^{ij}_h w_h^*, \; i = T, K, L, j = 1, \; 2, \tag{8}$$

where

$$\varepsilon^{ij}_h = \partial log a_{ij} / \partial log \; w_h = \theta_{hj} \sigma^{ij}_h. \tag{9}$$

$\sigma^{ij}_h$ is the Allen-partial elasticities of substitution (hereinafter AES) between the $i$th and the $h$th factors in the $j$th industry. For additional definition of these symbols, see Sato and Koizumi (1973, p.47-49), and BC (p.24). AESs are symmetric in the sense that

$$\sigma^{ij}_h = \sigma^{hj}_i. \tag{10}$$

And according to BC (p33), 'Given the assumption that production functions are strictly quasi-concave and linearly homogeneous,'

$$\sigma^{ij}_i < 0. \tag{11}$$

Since $a_{ij}$ is homogeneous of degree zero in input prices, we have

$$\Sigma_h \varepsilon^{ij}_h = \Sigma_h \theta_{hj} \sigma^{ij}_h = 0, \; i = T, K, L, j = 1, \; 2. \tag{12}$$

Equations (8) and (12) are equivalent to the expressions in BC (p.24, n. 6). See also JE (p.74, eqs (12)-(13)).

Substitute eq. (8) in (4):

$$\Sigma_j (\lambda_{ij} \Sigma_h \varepsilon^{ij}_h w_h^* + \lambda_{ij} X_j^*) = \Sigma_h g_{ih} w_h^* + \Sigma_j \lambda_{ij} X_j^* = V_i^*, \; i = T, K, L, \tag{13}$$

where

$$g_{ih} = \Sigma_j \lambda_{ij} \varepsilon^{ij}_h, i, h = T, K, L. \tag{14}$$

This is the EWS (or 'the economy-wide substitution') between factors $i$ and $h$ defined by JE (p.75). $g_{ih}$ is the aggregate of $\varepsilon^{ij}_{h}$. JE (p.75) stated, 'Clearly, the substitution terms in the two industries are always averaged together. With this in mind we define the term $\sigma^{i}_{k}$ to denote the economy-wide substitution towards or away from the use of factor $i$ when the kth factor becomes more expensive, under the assumption that each industry's output is kept constant:…'.

Note that

$$\Sigma_h g_{ih} = 0, i = T, K, L, \tag{15}$$
$$g_{ih} = (\theta_h / \theta_i) g_{hi}, i, h = T, K, L. \tag{16}$$

where $\theta_i$ and $\theta_j$ are, respectively, the share of factor $i$, $i = T, K, L$, and good $j$, $j = 1, 2$ in total income. That is, $\theta_j = p_j X_j / I$, $\theta_i = w_i V_i / I$, where $I = \Sigma_j p_j X_j = \Sigma_i w_i V_i$. On this, see BC (p.25, eq. (16)). Hence, we obtain $\lambda_{ij} = (\theta_j / \theta_i) \theta_{ij}$ (see JE (p.72, n. 9)). Note that $\Sigma_j \theta_j = 1$, $\Sigma_i \theta_i = 1$. $g_{ih}$ is not symmetric. Namely, $g_{ih} \neq g_{hi}, i \neq h$ in general. On eq. (16), see also JE (p.85).

From eqs (9), (11) and (14), we can show that

$$g_{ii} < 0. \tag{17}$$

From eqs (15) and (17), we derive

$$g_{KT} + g_{KL} = -g_{KK} > 0, g_{TK} + g_{TL} = -g_{TT} > 0, g_{LK} + g_{LT} = -g_{LL} > 0. \tag{18}$$

From eqs (18) and (16), we can easily show that

$$(g_{LK}, g_{LT}, g_{KT}) = (+,+,+), (-,+,+), (+,-,+), (+,+,-). \tag{19}$$

At most, one of EWSs $(g_{LK}, g_{LT}, g_{KT})$ can be negative.

2.2. Factor-intensity ranking

In this article, we assume

$$\theta_{T1}/\theta_{T2} > \theta_{L1}/\theta_{L2} > \theta_{K1}/\theta_{K2}, \tag{20}$$
$$\theta_{L1} > \theta_{L2}. \tag{21}$$

Equation (20) is, what you call, 'the factor-intensity ranking' (see JE (p69), see also BC (p.26-27), Suzuki (1983, p.142). This implies that sector 1 is relatively land intensive, sector 2 is relatively capital intensive, and that labor is the middle factor, and land and capital are extreme factors (see also Ruffin (1981, p.180)). Eq. (21) is 'the factor-intensity ranking for middle factor' (see JE (p. 70)). It implies that the middle factor is used relatively intensively in sector 1.

In the following sections, we show the analysis under these assumptions. Nakada (2017) also assumed eqs (20) and (21) hold. Of course, even if we assume $\theta_{L1} < \theta_{L2}$, we can analyze similarly.

2.3. EWS-ratio vector boundary

In this subsection, we derive the important relationship between EWS-ratios. I show the EWS-ratio vector boundary.

Each $a_{ij}$ is homogeneous of degree zero in all input prices (see eq. (3)). Recall eq. (11), $\sigma^{ij}_i < 0$. From these, Nakada (2017, eq. (35)) derived an important relationship among EWSs as follows.

$$g_{KK}g_{TT} - g_{TK}g_{KT} = g_{KT}g_{TL} + g_{KL}g_{TK} + g_{KL}g_{TL} = (\theta_L/\theta_T)[g_{KT}(g_{LT} + g_{LK}) + (\theta_L/\theta_K) g_{LK}g_{LT}] (>0). \quad (22)$$

Using eq. (18), transform eq. (22) to obtain

$$g_{KT} > -\frac{\theta_L}{\theta_K} \frac{g_{LK}g_{LT}}{g_{LK} + g_{LT}}. \quad (23)$$

Transform eq. (23) to have (see Nakada (2017))

$$U' > -\frac{\theta_L}{\theta_K} \frac{S'}{S'+1}, \text{ if T>0}; U' < -\frac{\theta_L}{\theta_K} \frac{S'}{S'+1}, \text{ if T<0}, \quad (24)$$

where

$$(S', U') = (S/T, U/T) = (g_{LK}/g_{LT}, g_{KT}/g_{LT}), \quad (25)$$
$$(S, T, U) = (g_{LK}, g_{LT}, g_{KT}). \quad (26)$$

We call $(S', U')$ the EWS-ratio vector. Equation (24) expresses the region for the EWS-ratio vector. Transform

$$U' = -\frac{\theta_L}{\theta_K}\frac{S'}{S'+1} = -\frac{\theta_L}{\theta_K} + \frac{\theta_L}{\theta_K}\frac{1}{S'+1},  \tag{27}$$

which expresses the rectangular hyperbola. We call it the equation for the EWS-ratio vector boundary. It passes on the origin of O (0, 0). The asymptotic lines are $S' = -1$, $U' = -\theta_L/\theta_K$. We can draw this boundary in the figure (see Fig. 1). S' is written along the horizontal axis, and U' along the vertical axis. This boundary demarcates the boundary of the region for the EWS-ratio vector. This implies that the EWS-ratio vector is not so arbitrary, but exists within this bounds.

The sign pattern of the EWS-ratio vector is, in each quadrant (see also eq. (19)):

Quad. I: $(S', U') = (+,+) \leftrightarrow (g_{LK}, g_{LT}, g_{KT}) = (+,+,+)$;
Quad. II: $(S', U') = (-,+) \leftrightarrow (g_{LK}, g_{LT}, g_{KT}) = (-,+,+)$;
Quad. III: $(S', U') = (-,-) \leftrightarrow (g_{LK}, g_{LT}, g_{KT}) = (+,-,+)$;
Quad. IV: $(S', U') = (+,-) \leftrightarrow (g_{LK}, g_{LT}, g_{KT}) = (+,+,-)$.  (28)

We may define (for $i \neq h$),

Factors $i$ and $h$ are economy-wide substitutes, if $g_{ih} > 0$;
Factors $i$ and $h$ are economy-wide complements, if $g_{ih} < 0$.  (29)

3. Estimating the position of the EWS-ratio vector
3.1. EWS-ratio vector line segment

In this subsection, we show that the EWS-ratio vector exists on the line segment. We have shown that the EWS-ratio vector exists on the straight line, which we call the EWS-ratio vector line (see eq. (B13) in Appendix B). The EWS-ratio vector satisfies eqs (B13) and (24). Using eqs (B13) and (27), make a system of equations:

$$U' = -a_1 S' + b_1,  \tag{30}$$

$$U' = -\frac{\theta_L}{\theta_K}\frac{S'}{S'+1}.  \tag{31}$$

From these, we obtain a quadratic equation in *S'* for each *i, j*. Solve this to derive two solutions. Each solution denotes the *S'* coordinate value of the intersection point of the EWS-ratio vector line and the EWS-ratio vector boundary. The solutions are[5]

$$S' = \frac{-W_{TL}}{W_{KL}}, \frac{a_{K0}'}{a_{T0}'}\theta_{KT}, \qquad (32)$$

Hence, the Cartesian coordinates of the intersection point are

$$(S',U') = (\frac{-W_{TL}}{W_{KL}}, \frac{\theta_L}{\theta_K}\frac{-W_{LT}}{W_{KT}}), (\frac{a_{K0}'}{a_{T0}'}\theta_{KT}, \frac{a_{K0}'}{a_{L0}'}). \qquad (33)$$

We call these points *A* and *B*.

In general, the EWS-ratio vector $(S',U')$ exists on the line segment *AB*. We call it the EWS-ratio vector line segment (see Fig.1). This implies that by analyzing the Cartesian coordinates of points *A* and *B*, we can estimate the position of the EWS-ratio vector.

3.2. Factor-price-change ranking and its implications

The changes in the real factor price are not independent but are related to each other. They satisfy the factor-price-change ranking. For ease of notation, define that

$$(X,Y,Z) = (w_{T1}{}^*, w_{K1}{}^*, w_{L1}{}^*) = (w_T{}^* - p_1{}^*, w_K{}^* - p_1{}^*, w_L{}^* - p_1{}^*). \qquad (34)$$

The following result has been established.

Lemma 1 We assume the factor intensity ranking as in eqs (20) and (21), and the change in the relative price of goods as follows.

$$P = p_1{}^* - p_2{}^* > 0. \qquad (35)$$

This implies that only four factor-price-change rankings are possible, that is,

---

[5] For the definition of symbols, recall eqs (B14), (B2), and (B8), that is,

$W_{ih} = w_i{}^* - w_h{}^* = (w_i / w_h)^*, i = T, K, L, i \neq h,$  $a_{i0}' = \sum_j \lambda_{ij} a_{ij}{}^*, i = T, K, L,$

$\theta_{ih} = \theta_i / \theta_h, i = T, K, L, i \neq h.$

$$X > Y > Z, \ X > Z > Y, \ Z > X > Y, \ Z > Y > X. \tag{36}$$

Proof. See Appendix C.

The following result has been established.

Lemma 2 We assume the factor intensity ranking and the change in the relative price of goods as follows.

$$\theta_{T1}/\theta_{T2} > \theta_{L1}/\theta_{L2} > \theta_{K1}/\theta_{K2}, \ \theta_{L1} > \theta_{L2}, \tag{37}$$
$$P = p_1{}^* - p_2{}^* > 0. \tag{38}$$

And, further, if we assume the factor-price-change ranking as follows (from Lemma 1, this assumption is plausible enough)

$$X > Z > Y \leftrightarrow w_T{}^* > w_L{}^* > w_K{}^*, \tag{39}$$

the signs A, B, C, D are possible. That is,

$$\begin{matrix} & A & B & C & D \end{matrix}$$
$$(a_{T0}', a_{K0}', a_{L0}') = (-,+,-),(-,+,+),(+,+,-),(-,-,+), \tag{40}$$
$$(a_{Tj}{}^*, a_{Kj}{}^*, a_{Lj}{}^*) = (-,+,-),(-,+,+),(+,+,-),(-,-,+), \ j=1.2, \tag{41}$$

where $a_{i0}' = \sum_j \lambda_{ij} a_{ij}{}^*, i = T, K, L$. $a_{i0}'$ is the aggregate of $a_{ij}{}^*$ (or the rate of change in the input-output coefficient).

Proof. See Appendix D.

3.3. Estimating the position of the EWS-ratio vector

In this subsection, we assume eqs (37), (38), and (39) in Lemma 2 hold. We estimate the position of the EWS-ratio vector. In subsection 3.3.1, we derive a sufficient condition for the EWS-ratio vector to exist in quadrant IV. Additionally, in subsection 3.3.2, we derive a sufficient condition for the EWS-ratio vector to exist in a specific subregion. In subsection 3.3.3, I show some implications.

3.3.1. A sufficient condition for the EWS-ratio vector to exist in quadrant IV

The following result has been established.[6]

Theorem 1 We assume the factor intensity ranking and the change in the relative price of goods as follows.

$$\theta_{T1}/\theta_{T2} > \theta_{L1}/\theta_{L2} > \theta_{K1}/\theta_{K2}, \quad \theta_{L1} > \theta_{L2}, \tag{42}$$

$$P = p_1{}^* - p_2{}^* > 0. \tag{43}$$

The EWS-ratio vector $(S', U')$ exists on the EWS-ratio vector line segment (or the line segment AB). Using this relationship, we can estimate the position of the EWS-ratio vector. For example, if we assume (from Lemma 2, these assumptions are plausible enough)

$$X > Z > Y \leftrightarrow w_T{}^* > w_L{}^* > w_K{}^*, \text{ and } (a_{T0}', a_{K0}', a_{L0}') = (+, +, -), \tag{44}$$

the Cartesian coordinates of points A and B are, respectively:

$$\left(\frac{-W_{TL}}{W_{KL}}, \frac{\theta_L}{\theta_K}\frac{-W_{LT}}{W_{KT}}\right) = \left(\frac{-(w_T{}^* - w_L{}^*)}{(w_K{}^* - w_L{}^*)}, \frac{\theta_L}{\theta_K}\frac{-(w_L{}^* - w_T{}^*)}{(w_K{}^* - w_T{}^*)}\right) = (+, -), \tag{45}$$

$$\left(\frac{a_{K0}'}{a_{T0}'}\theta_{KT}, \frac{a_{K0}'}{a_{L0}'}\right) = (+, -). \tag{46}$$

Hence, both of points A and B are in quadrant IV, and, point A is on the left-hand side of point B. The line segment AB exists in quadrant IV. Hence, the EWS-ratio vector is in quadrant IV and satisfies

$$0 < \frac{-W_{TL}}{W_{KL}} < S' < \frac{a_{K0}'}{a_{T0}'}\theta_{KT}, \quad 0 > \frac{\theta_L}{\theta_K}\frac{-W_{LT}}{W_{KT}} > U' > \frac{a_{K0}'}{a_{L0}'}. \tag{47}$$

---

[6] From Lemma 2, for $a_{i0}'$, the signs A, B, C, D are possible (see (D17)). Therefore, for point B, if signs A, B, C, and D hold, we derive, respectively:

$$\begin{array}{cccc} \text{A} & \text{B} & \text{C} & \text{D} \end{array}$$
$$\left(\frac{a_{K0}'}{a_{T0}'}\theta_{KT}, \frac{a_{K0}'}{a_{L0}'}\right) = (-, -), (-, +), (+, -), (+, -).$$

Hence, point B is in quadrant III, II, IV, and IV, respectively. If sign D holds, both points A and B are in quadrant IV. We can easily show that point A is on the right-hand side of point B. However, we omit this analysis.

In this case, capital and land, extreme factors, are economy-wide complements. Hence, a strong Rybczynski result holds, that is, three of the Rybczynski sign patterns hold (see Theorem A.1).
Proof. See Appendix E.

We have derived a sufficient condition for extreme factors to be economy-wide complements. Equation (44) implies the following. The rate of change in real reward for labor is intermediate (or moderate), and the rate of change in real reward for land and capital are extreme. Both of the signs of the aggregate of $a_{Tj}*$ and $a_{Kj}*$ (or the rate of change in the input–output coefficients of land and capital, respectively) are positive, and the sign of the aggregate of $a_{Lj}*$ (or the rate of change in the input–output coefficient of labor) is negative.

3.3.2. A sufficient condition for the EWS-ratio vector to exist in a specific subregion

In this subsection, we assume eq. (44) holds, hence, eqs (47) holds.

By comparing the $S'$ coordinates of Points $R_{L2}$ and $R_{L1}$ in Fig. A1 with the $S'$ coordinates of points $A$ and $B$ in Fig. 1, we can show a sufficient condition for the EWS-ratio vector to exist in a specific subregion. The Cartesian coordinates of Points $R_{L2}$ and $R_{L1}$ are, respectively (Nakada (2017, eq. (71)):

$$(S', U') = (\frac{\theta_{K1}}{\theta_{T1}}, \frac{-\theta_{K1}}{1-\theta_{L1}} \frac{\theta_L}{\theta_K}), (\frac{\theta_{K2}}{\theta_{T2}}, \frac{-\theta_{K2}}{1-\theta_{L2}} \frac{\theta_L}{\theta_K}). \tag{48}$$

For the relationship between the Cartesian coordinates of points $A$ and $B$, and the EWS-ratio vector $(S', U')$, see eq. (47). Using eqs (47) and (48), we derive the following result. A corollary of Theorem 1 is as follows.

Corollary 1 We assume eqs (42)-(44) in Theorem 1 hold.
(i) Example 1: If the equation shown below holds,

$$\frac{\theta_{K2}}{\theta_{T2}} < \frac{-W_{TL}}{W_{KL}} < S' < \frac{a_{K0}'}{a_{T0}'} \theta_{KT}, \tag{49}$$

both points $A$ and $B$ on the EWS-ratio vector line segment exist in subregion P1. Hence, the EWS-ratio vector exists in subregion P1. The sufficient condition for eq. (49) is,

$$\frac{\theta_{K2}}{\theta_{T2}} < \frac{-W_{TL}}{W_{KL}} \leftrightarrow w_L* - p_2* = (-) < 0. \tag{50}$$

(ii) Example 2: If the equation shown below holds,

$$\frac{\theta_{K1}}{\theta_{T1}} < \frac{-W_{TL}}{W_{KL}} < S' < \frac{a_{K0}'}{a_{T0}'}\theta_{KT} < \frac{\theta_{K2}}{\theta_{T2}}, \tag{51}$$

both points *A* and *B* exist in subregion P2. Hence, the EWS-ratio vector exists in subregion P2. The sufficient condition for eq. (51) is the set of equations shown below.

$$\frac{-W_{TL}}{W_{KL}} < \frac{\theta_{K2}}{\theta_{T2}} \leftrightarrow w_L{}^* - p_2{}^* = (+) > 0, \tag{52}$$

$$\frac{\theta_{K1}}{\theta_{T1}} < \frac{-W_{TL}}{W_{KL}} \leftrightarrow w_L{}^* - p_1{}^* = (-) < 0, \text{ and} \tag{53}$$

$$\frac{a_{K0}'}{a_{T0}'}\frac{\theta_K}{\theta_T} < \frac{\theta_{K2}}{\theta_{T2}}. \tag{54}$$

(iii) Example 3: If the equation shown below holds,

$$\frac{-W_{TL}}{W_{KL}} < S' < \frac{a_{K0}'}{a_{T0}'}\theta_{KT} < \frac{\theta_{K1}}{\theta_{T1}}, \tag{55}$$

both points *A* and *B* exist in subregion P3. Hence, the EWS-ratio vector exists in subregion P3. The sufficient condition for eq. (55) is the set of equations shown below.

$$\frac{-W_{TL}}{W_{KL}} < \frac{\theta_{K1}}{\theta_{T1}} \leftrightarrow w_L{}^* - p_1{}^* = (+) > 0, \tag{56}$$

$$\frac{a_{K0}'}{a_{T0}'}\frac{\theta_K}{\theta_T} < \frac{\theta_{K1}}{\theta_{T1}}. \tag{57}$$

Proof. For example, we show the proof of eq. (50). Multiplying eq. (50) by $W_{KL}(= w_K{}^* - w_L{}^* < 0)$, we have

$$\theta_{K2}W_{KL} > -W_{TL}\theta_{T2} \leftrightarrow \theta_{K2}(w_K{}^* - w_L{}^*) > -(w_T{}^* - w_L{}^*)\theta_{T2}$$

$$\leftrightarrow \sum_i \theta_{i2} w_i{}^* - w_L{}^* > 0. \tag{58}$$

Using eq. (6), transform (58) to derive

$$p_2{}^* - w_L{}^* > 0 \leftrightarrow w_L{}^* - p_2{}^* = (-) < 0. \tag{59}$$

We omit the proof of other equations.

### 3.3.3. Some implications

We show some implications of Theorem 1. We assume eqs (42)-(44) hold. Hence, extreme factors must be economy-wide complements. For example, Thompson and Ban assumed as follows, respectively.

(i) Thompson (1995) assumed that production functions are of a Cobb-Douglas type and an all constant CES type.
(ii) Ban (2007a, 2008) assumed that production functions are of the two-level CES type. See also Ban (2007b).[7]

However, are these studies plausible?[8]

We can show that it is implausible to assume as in (i). It is because they do not allow any two factors to be Allen-complements. Moreover, we can show that it is questionable to assume as in (ii). I show the proof in Appendix F.

### 4. Conclusion

---

[7] In her model, three factors are skilled labor, capital, and unskilled labor. Ban (2007a) assumed that skilled labor and capital could be '[Allen-] complements' in each sector, and she computed the values of AESs theoretically. Ban (2007a) attempted to analyze how commodity prices affect relative factor prices. She described the effects when she changed factor-intensity ranking. However, her analysis is somewhat complicated, and her results are not clear. This is a theoretical study. Ban (2008, p. 4, Table 1) showed a table classifying the results in Ban (2007a) by factor-intensity ranking. She classified the countries in the world into 14 regions in total and computed the factor-intensity for each area using the GTAP version 6 database. Additionally, she assumed 10 types of values for 'the elasticities of substitution' (equivalent to EWS) to simulate how commodity prices affect the relative factor prices. This is an application. Ban (2011, chapter 4, p. 87–89) summarized Ban (2007a) and Ban (2008), and modified the studies. For her results, see Ban (2011, p. 96–97, Table 4–1). On this, see also Nakada (2017).

[8] Of course, in normal CGE (or computable general equilibrium) analysis, it is usual to assume Cobb-Douglas type or CES type. On this, for example, see Bergman (2005. P. 1285–6).

In this article, we have assumed a certain pattern of factor-intensity ranking including a certain pattern of factor-intensity ranking of the middle factor (On this, see eqs (20) and (21)). That is, we have assumed that sector 1 is relatively land intensive, sector 2 is relatively capital intensive, and that labor is the middle factor, and land and capital are extreme factors. Further, we have assumed that the middle factor is used relatively intensively in sector 1.

In the Introduction, I posed the following questions.

(i)     How can we estimate the position of the EWS-ratio vector?

(ii)    Under what conditions does the EWS-ratio vector exist in quadrant IV; in other words, are extreme factors economy-wide complements?

We derive the results as follows.

Answer to (i): We have shown that the EWS-ratio vector exists on the line segment *AB* (or the EWS-ratio vector line segment). Points *A* and *B* are the intersection points of the EWS-ratio vector line and the EWS-ratio vector boundary (see Fig. 1). Using this relationship, we have developed a method to estimate the position of the EWS-ratio vector. That is, with the appropriate data, we know the position of points *A* and *B*, hence, we also know the position of the EWS-ratio vector to some extent.

Answer to (ii): First, we have derived a sufficient condition for the EWS-ratio vector to exist in quadrant IV (that is, in subregion P1, P2, or P3) (see Theorem 1). If this holds, from Theorem A.1, a strong Rybczynski result holds necessarily, that is, three of the Rybczynski sign patterns hold. We call this a rough estimate.

Additionally, we have derived a sufficient condition for the EWS-ratio vector to exist in a specific subregion. If this holds, a specific Rybczynski sign pattern holds (see Corollary 1). If we use this property, we can conduct a detailed estimate.

To derive the sufficient condition shown in Theorem 1, we need data on the change in some variables, which requires data for two time points. That is, the sign of the change in the relative price of a commodity, the factor-price-change ranking, and the sign of $a_{i0}'$, or the aggregate of $a_{ij}*$ (or the rate of change in the input–output coefficient). On the other hand, normal CGE (or computable general equilibrium) analysis only requires the data for one time-point to estimate the value of basic parameters. To do a detailed estimate, we need more detailed data.

This article suggests the following. In some cases, it is implausible to assume that production functions are of a Cobb-Douglas type, or an all-constant CES type in each sector, which do not allow any two factors to be Allen-complements. Moreover, it is questionable to assume that production functions are of the two-level CES type.

This article provides a basis for further applications. For example, this article contributes to the estimation of the Rybczynski sign pattern in some countries and contributes to international and development economics.[9]

It is uncertain whether we can reduce the range of the EWS-ratio vector further.

Equation Section (Next)

Appendix A: Analysis in Nakada (2017)

I briefly explain the analysis in Nakada (2017). Nakada (2017) drew the border line for a Rybczynski sign pattern to change in the figure, which we call line *ij*. Line *ij* divides the region of the EWS-ratio vector into 12 subregions (P1-7, M1-5) (see Fig. A1). There are seven intersection points of line *ij* and the EWS-ratio vector boundary. Each line *ij* passes through the same point, which we call point Q. The Cartesian coordinates of point Q are

$$(S', U') = (B/A, (B/E)(\theta_L/\theta_K)) = (-,-), \tag{A1}$$

where $(A, B, E) = (\theta_{T1} - \theta_{T2}, \theta_{K1} - \theta_{K2}, \theta_{L1} - \theta_{L2})$. We call six intersection points other than point Q the point $R_{ij}, i = T, K, L, j = 1, 2$. I omit the Cartesian coordinates of these points.

Nakada (2017) concluded that the position of the EWS-ratio vector determines the Rybczynski sign pattern. Nakada derived a sufficient condition for a strong Rybczynski result to hold (or not to hold). That is, the following result has been established as mentioned in the Introduction (see Nakada (2017, Theorem 1)). We have rearranged it below.

*Theorem A.1* We assume the factor-intensity ranking as follows.

$$\theta_{T1}/\theta_{T2} > \theta_{L1}/\theta_{L2} > \theta_{K1}/\theta_{K2}, \tag{A2}$$

$$\theta_{L1} > \theta_{L2}. \tag{A3}$$

Further, if the EWS-ratio vector $(S', U')$ exists in quadrant IV (or subregions P1-P3), in other words, if extreme factors are economy-wide complements, a strong Rybczynski result necessarily holds. In this case, the Rybczynski sign patterns, as per Thompson's (1985, p. 619) terminology, for subregions P1-P3 are, respectively:

P1      P2      P3

---

[9] For example, Nakada (2016b) applied the results derived here to data from Thailand and, in doing so, derived the factor endowment–commodity output relationship for Thailand during the period 1920–29. To some extent, these results show how Chinese immigration affected commodity output in Thailand between 1920 and 1929.

$$sign[X_j*/V_i*] = \begin{bmatrix} + & - & - \\ - & + & + \end{bmatrix} \begin{bmatrix} + & - & + \\ - & + & + \end{bmatrix} \begin{bmatrix} + & - & + \\ - & + & - \end{bmatrix}, \quad (A4)$$

where

$$[X_j*/V_i*] = \begin{bmatrix} X_1*/V_T* & X_1*/V_K* & X_1*/V_L* \\ X_2*/V_T* & X_2*/V_K* & X_2*/V_L* \end{bmatrix}. \quad (A5)$$

Equations (A4) expresses the factor endowment–commodity output relationships.

For the Stolper-Samuelson sign patterns, see Nakada (2017).

Equation Section (Next)

Appendix B: The EWS-ratio vector line

In this appendix, we derive a linear relationship between EWS-ratios. Using eq. (12), eliminate $\varepsilon^{ij}{}_i$ from eq. (8) to obtain[10]

$$a_{Tj}* = \varepsilon_{TKj}(w_K*-w_T*) + \varepsilon_{TLj}(w_L*-w_T*),$$
$$a_{Kj}* = \varepsilon_{KLj}(w_L*-w_K*) + \varepsilon_{KTj}(w_T*-w_K*),$$
$$a_{Lj}* = \varepsilon_{LTj}(w_T*-w_L*) + \varepsilon_{LKj}(w_K*-w_L*), \quad (B1)$$

where $\varepsilon_{ihj} = \varepsilon^{ij}{}_h$.

Define that

$$a_{i0}' = \sum_j \lambda_{ij} a_{ij}*, i = T, K, L. \quad (B2)$$

This is the aggregate of $a_{ij}*$. Substitute eq. (B1) in (B2). Rewrite using eq. (14) to have

$$a_{T0}' = \sum_j \lambda_{Tj}\{\varepsilon^{Tj}{}_K(w_K*-w_T*) + \varepsilon^{Tj}{}_L(w_L*-w_T*)\}$$
$$= g_{TK}(w_K*-w_T*) + g_{TL}(w_L*-w_T*) \quad (B3)$$

Similarly, we derive

$$a_{K0}' = g_{KL}(w_L*-w_K*) + g_{KT}(w_T*-w_K*), \quad (B4)$$
$$a_{L0}' = g_{LT}(w_T*-w_L*) + g_{LK}(w_K*-w_L*). \quad (B5)$$

Using eq. (16), eliminate $g_{TK}$, $g_{TL}$, and $g_{KL}$ from eqs (B3) and (B4) to obtain

---

[10] Equation (B1) is equivalent to eqs (10)-(12) in BC (p. 24), who used AESs.

$$a_{T0}' = \theta_{KT}g_{KT}(w_K^* - w_T^*) + \theta_{LT}g_{LT}(w_L^* - w_T^*),\tag{B6}$$

$$a_{K0}' = \theta_{LK}g_{LK}(w_L^* - w_K^*) + g_{KT}(w_T^* - w_K^*),\tag{B7}$$

Where

$$\theta_{ih} = \theta_i / \theta_h, i \neq h.\tag{B8}$$

Multiply eqs (B6), (B7), and (B5) by $g_{LK}\theta_T$, $g_{LT}\theta_K$, and $g_{KT}\theta_K$, respectively, and take the difference to obtain

$$a_{T0}'g_{LK}\theta_T - a_{K0}'g_{LT}\theta_K = (w_K^* - w_T^*)G_0,$$
$$a_{K0}'g_{LT}\theta_K - a_{L0}'g_{KT}\theta_K = (w_L^* - w_K^*)G_0,$$
$$a_{L0}'g_{KT}\theta_K - a_{T0}'g_{LK}\theta_T = (w_T^* - w_L^*)G_0,\tag{B9}$$

where

$$G_0 = g_{KT}\theta_K(g_{LK} + g_{LT}) + g_{LT}g_{LK}\theta_L > 0.\tag{B10}$$

Equation (B10) is derived from eq. (22). From eq. (B9), we have

$$G_0 = \frac{a_{T0}'g_{LK}\theta_T - a_{K0}'g_{LT}\theta_K}{w_K^* - w_T^*} = \frac{a_{K0}'g_{LT}\theta_K - a_{L0}'g_{KT}\theta_K}{w_L^* - w_K^*}.\tag{B11}$$

Divide the both sides of eq. (B11) by $g_{LT}$, and transform using EWS-ratios to have (see eq (25)),

$$(a_{T0}'S'\theta_T - a_{K0}'\theta_K)(w_L^* - w_K^*) = (a_{K0}'\theta_K - a_{L0}'U'\theta_K)(w_K^* - w_T^*).\tag{B12}$$

From eq. (B12), we derive

$$U' = -a_1 S' + b_1,\tag{B13}$$

where

$$a_1 = \frac{a_{T0}'\theta_T W_{LK}}{a_{L0}'\theta_K W_{KT}}, b_1 = \frac{a_{K0}'W_{LT}}{a_{L0}'W_{KT}}, W_{ih} = w_i^* - w_h^* = (w_i / w_h)^*, i, h = T, K, L, i \neq h.\tag{B14}$$

$W_{ih}$ is the change in relative factor price between factors *i* and *h*. Equation (B13) expresses the straight line, which we call the EWS-ratio vector line. The EWS-ratio vector $(S', U')$ exists on this line. Hence, *U'* has a linear relationship with *S'*.

Equation Section (Next)

Appendix C: Proof of Lemma 1

In this appendix, we show the relationship among the factor-price changes. Using eq. (34), rewrite eq. (7). Transform to derive

$$\begin{bmatrix} \theta_{T1} & \theta_{K1} \\ \theta_{T2} & \theta_{K2} \end{bmatrix} \begin{bmatrix} X \\ Y \end{bmatrix} = \begin{bmatrix} -\theta_{L1} Z \\ -P - \theta_{L2} Z \end{bmatrix}. \tag{C1}$$

Solving eq. (C1), we have

$$X = \theta_{K1} P / D_1 - (D_2 / D_1) Z,$$
$$Y = -\theta_{T1} P / D_1 - (D_3 / D_1) Z, \text{ and}$$
$$Z = Z, \tag{C2}$$

where $D_1 = \theta_{T1}\theta_{K2} - \theta_{K1}\theta_{T2}, D_2 = \theta_{K2}\theta_{L1} - \theta_{K1}\theta_{L2}, D_3 = \theta_{T1}\theta_{L2} - \theta_{T2}\theta_{L1}$. Equation (20) implies

$$(D_1, D_2, D_3) = (+, +, +). \tag{C3}$$

Treat as if *X* and *Y* were dependent variables and *Z* were an independent variable. Equation (C2) expresses the straight lines in two dimensions. We call these lines, respectively, Lines *X*, *Y*, and *Z*. From eqs (C3) and (35), the signs of the gradient and intercept of Lines *X* and *Y* are, respectively,

$$-(D_2/D_1) = (-), \ \theta_{K1} P / D_1 = (+), \text{ for Line } X;$$
$$-(D_3/D_1) = (-), \ -\theta_{T1} P / D_1 = (-), \text{ for Line } Y. \tag{C4}$$

Because eq. (21) holds, if we compare the gradient of Lines *X* and *Y*, we can easily show that

$$-(D_2/D_1) < -(D_3/D_1). \tag{C5}$$

Hence, we can draw Lines *X, Y,* and *Z* in the figure (see Fig**.** C1). Lines *X* and *Y* have an intersection point in quadrant IV. The S' coordinates of the intersection of Lines *Y* and *Z* and Lines *X* and *Z* are, respectively,

$$[-\theta_{T1}/(\theta_{T1} - \theta_{T2})]P, [-\theta_{K1}/(\theta_{K1} - \theta_{K2})]P. \tag{C6}$$

Only four rankings are possible; they are,

$$X > Y > Z, \ X > Z > Y, \ Z > X > Y, \ Z > Y > X. \tag{C7}$$

We call this the factor-price-change ranking.

Equation Section (Next)

Appendix D: Proof of Lemma 2

First, we derive the important relationships among some variables ($\Delta w_i$ and $\Delta a_{ij}$). Next, using these relationships, we prove Lemma 2. Equation (2) expresses the isocost surface. We can draw the isocost surface (or IC) and the isoquant surface (or IQ) in the figure (see Fig. D1). Define that

$$\vec{w} = w_i, \vec{w}' = w_i + \Delta w_i, \overrightarrow{OA} = a_{ij}, \overrightarrow{OB} = a_{ij} + \Delta a_{ij}, \overrightarrow{AB} = \Delta a_{ij}, i = T, K, L, \ j = 1, 2, \tag{D1}$$

where $\Delta$ denotes the small variation. Vector $w_i$ is vertical to the isocost surface. Because production functions are homogeneous of degree one and strictly quasi-concave, the isoquant surface is convex to the origin. The isoquant surface touches the isocost surface at point $A$. That point is the equilibrium point. $a_{ij}$ (or the input-output coefficient) is determined by this point. We draw this figure by analogy from the figure of the isocost line and the isoquant curve for the two-factor case. If the isocost surface changes its position and becomes IC', the equilibrium point moves to point $B$. Angles $\theta_A$ and $\theta_B$ are the angles between vectors $\vec{w}$ and $\overrightarrow{AB}$, $\vec{w}'$ and $\overrightarrow{BA}$, respectively.

We obtain

$$-\pi/2 < \theta_A < 0, 0 < \theta_B < \pi/2. \tag{D2}$$

Hence, the inner product of vectors satisfies

$$\vec{w} \cdot \overrightarrow{AB} = \sum_i w_i \Delta a_{ij} = |\vec{w}||\overrightarrow{AB}| \cos \theta_A > 0, \tag{D3}$$

$$\vec{w}' \cdot \overrightarrow{BA} = \sum_i (w_i + \Delta w_i)(-\Delta a_{ij}) = |\vec{w}'||\overrightarrow{BA}| \cos \theta_B > 0. \tag{D4}$$

Sum up eqs (D3) and (D4), and multiply the both sides by -1 to have[11]

$$H_j = \sum_i \Delta w_i \Delta a_{ij} < 0, \ j = 1, \ 2. \tag{D5}$$

Transforming eq. (D5), we can express in elasticity terms:

$$H_j = \sum_i w_i * a_{ij} * \theta_{ij} p_j < 0, \ j = 1, \ 2. \tag{D6}$$

Define that

$$H_0 = \sum_j \theta_j (H_j / p_j). \tag{D7}$$

We call this the aggregate of $H_j / p_j$. Substituting eq. (D6) in (D7), we have

$$H_0 = \sum_i \sum_j w_i * \lambda_{ij} a_{ij} * \theta_i < 0. \tag{D8}$$

Using eq. (B2), rewrite (D8) to obtain

$$H_0 = \sum_i w_i * a_{i0}' \theta_i < 0. \tag{D9}$$

From eqs (B2) and (5), we can show that

$$\sum_i a_{i0}' \theta_i = 0. \tag{D10}$$

Equation (D10) implies that

$$(a_{T0}', a_{K0}', a_{L0}') = (-,+,-),(-,+,+),(+,+,-),(-,-,+),(+,-,+),(+,-,-). \tag{D11}$$

One or two of $a_{i0}'$ must be negative. We call these sign patterns the signs A, B, C, D, E, and F, respectively.

Using eq. (D10), eliminate $a_{L0}' \theta_L$ from (D9) to derive

$$H_0 = (w_T * - w_L *) a_{T0}' \theta_T + (w_K * - w_L *) a_{K0}' \theta_K < 0. \tag{D12}$$

Similarly, we derive

---

[11] Equation (D5) is very similar to the equation which Samuelson (1983, Chapter 4, p. 78, eq. (82)) derived, that is, $\sum_1^n \Delta w_i \Delta v_i \leq 0$, where $w_i$ is the price of factor $i$; $v_i$ is the combination of factors that minimize the total cost, but the author derived this equation differently.

$$H_0 = (w_T{*}-w_K{*})a_{T0}'\theta_T + (w_L{*}-w_K{*})a_{L0}'\theta_L < 0, \tag{D13}$$

$$H_0 = (w_K{*}-w_T{*})a_{K0}'\theta_K + (w_L{*}-w_T{*})a_{L0}'\theta_L < 0. \tag{D14}$$

Equations (D12)-(D14) are just like constraints on some variables ($w_i{*}-w_h{*}$ and $a_{i0}'$). Next, we show the implications of eqs (D12)-(D14).

For example, we assume eqs (38) and (39) in Lemma 2 hold. From eq. (39), we derive

$$(w_T{*}-w_L{*}, w_K{*}-w_L{*}) = (+,-), (w_T{*}-w_K{*}, w_L{*}-w_K{*}) = (+,+),$$
$$(w_K{*}-w_T{*}, w_L{*}-w_T{*}) = (-,-). \tag{D15}$$

Substituting eq. (D15) in (D12)-(D14), we obtain

$$(a_{T0}', a_{K0}', a_{L0}') \neq (+,-,+), (+,-,-). \tag{D16}$$

Hence, the signs E and F are impossible. The signs A, B, C, and D are possible (see eq. (D11)). That is,

$$\phantom{(a_{T0}', a_{K0}', a_{L0}') = }\ \ \text{A}\phantom{xxxx}\text{B}\phantom{xxxx}\text{C}\phantom{xxxx}\text{D}$$
$$(a_{T0}', a_{K0}', a_{L0}') = (-,+,-), (-,+,+), (+,+,-), (-,-,+). \tag{D17}$$

Similarly, if we assume (D15), using eqs (5) and (D6), we can show that the signs A, B, C, and D are possible for sector $j$. That is,

$$\phantom{(a_{Tj}{*}, a_{Kj}{*}, a_{Lj}{*}) = }\ \ \text{A}\phantom{xxxx}\text{B}\phantom{xxxx}\text{C}\phantom{xxxx}\text{D}$$
$$(a_{Tj}{*}, a_{Kj}{*}, a_{Lj}{*}) = (-,+,-), (-,+,+), (+,+,-), (-,-,+), j=1.2. \tag{D18}$$

On this, see Nakada (2016a, p. 11-13).

<span style="color:red">Equation Section (Next)</span>

Appendix E: Proof of Theorem 1

We can show which point is on the left-hand side, points *A* or *B*. Transforming eq. (D12), we have

$$H_0 = (w_K{*}-w_L{*})a_{T0}'\theta_T\{\frac{W_{TL}}{W_{KL}} + \frac{a_{K0}'}{a_{T0}'}\theta_{KT}\} < 0. \tag{E1}$$

From eqs (44), we have

$$w_K{*}-w_L{*} = (-), a_{T0}' = (+). \tag{E2}$$

From eqs (E1) and (E2), we derive

$$\frac{W_{TL}}{W_{KL}} + \frac{a_{K0}'}{a_{T0}'}\theta_{KT} > 0 \leftrightarrow -\frac{W_{TL}}{W_{KL}} < \frac{a_{K0}'}{a_{T0}'}\theta_{KT}. \tag{E3}$$

Hence, point *A* is on the left-hand side of point *B* (see Fig. 1).

From eq. (E3), the EWS-ratio vector $(S',U')$ satisfies

$$0 < \frac{-W_{TL}}{W_{KL}} < S' < \frac{a_{K0}'}{a_{T0}'}\theta_{KT}, 0 > \frac{\theta_L}{\theta_K}\frac{-W_{LT}}{W_{KT}} > U' > \frac{a_{K0}'}{a_{L0}'}. \tag{E4}$$

Equation Section (Next)

Appendix F: Investigation of the assumption that production functions are of the two-level CES type

In this appendix, we investigate the assumption that production functions are of the two-level CES type. We assume eqs (42)-(44) hold.

Replace skilled labor, capital, and unskilled labor (*S, K, L*) in Ban (2007a) with land, capital, and labor (*T, K, L*), respectively. Ban (2007a, 2008) changed factor-intensity ranking, but this seems confusing. We assume the factor-intensity ranking is constant as shown in eqs (20) and (21). And, we assume *T* and *K* could be [Allen-] complements. The values of AES are

$$(\sigma^{Kj}_L, \sigma^{Lj}_T, \sigma^{Tj}_K) = (c_j, c_j, \sigma^{Tj}_K), j = 1, 2, \tag{F1}$$

where $c_j$ is constant (>0); $\sigma^{Tj}_K$ is variable, $\sigma^{Tj}_K > 0$, or $\sigma^{Tj}_K < 0$. From eqs (26), (14), and (9), we derive

$$\begin{aligned}(S, T, U) &= (g_{LK}, g_{LT}, g_{KT}) = (\Sigma_j \lambda_{Lj}\varepsilon^{Lj}_K, \Sigma_j \lambda_{Lj}\varepsilon^{Lj}_T, \Sigma_j \lambda_{Kj}\varepsilon^{Kj}_T) \\ &= (\Sigma_j \lambda_{Lj}\theta_{Kj}\sigma^{Lj}_K, \Sigma_j \lambda_{Lj}\theta_{Tj}\sigma^{Lj}_T, \Sigma_j \lambda_{Kj}\theta_{Tj}\sigma^{Kj}_T).\end{aligned} \tag{F2}$$

The EWS contains AESs.

Substituting eq. (F1) in (F2), we derive

$$S > 0, T > 0, \text{ and } U > 0, \text{ or } U < 0, \tag{F3}$$

where *S* and *T* are constant. However, the value of *U* is variable. Substitute eq. (F3) in (25) to have

$$(S',U') = (d,?), \tag{F4}$$

where $d$ is constant ($>0$). Hence, $S'$ is constant ($>0$), and $U'$ is variable ($>0$, or $<0$). Hence, the EWS-ratio vector is on the line

$$S' = d . \tag{F5}$$

However, is this plausible? In this article, we have shown that the EWS-ratio vector exists on the line segment $AB$ (see eq. (33)).

We assume eqs (42)-(44) hold as mentioned earlier. Therefore, the EWS-ratio vector is in quadrant IV and satisfies eq. (47). The line segment $AB$ exits in quadrant IV (see Fig. 1). And it might

(i)  join the line $S' = d$, or
(ii) not join.

In case of (i), it joins only at one point. Ban (2008) assumed 10 types of values for 'the elasticities of substitution' (equivalent to EWS) to simulate how commodity prices affect the relative factor prices. However, only one value might be plausible. In case of (ii), the value of $(S', U')$ on the line $S' = d$ is not possible at all. Apparently, this simulation seems questionable.

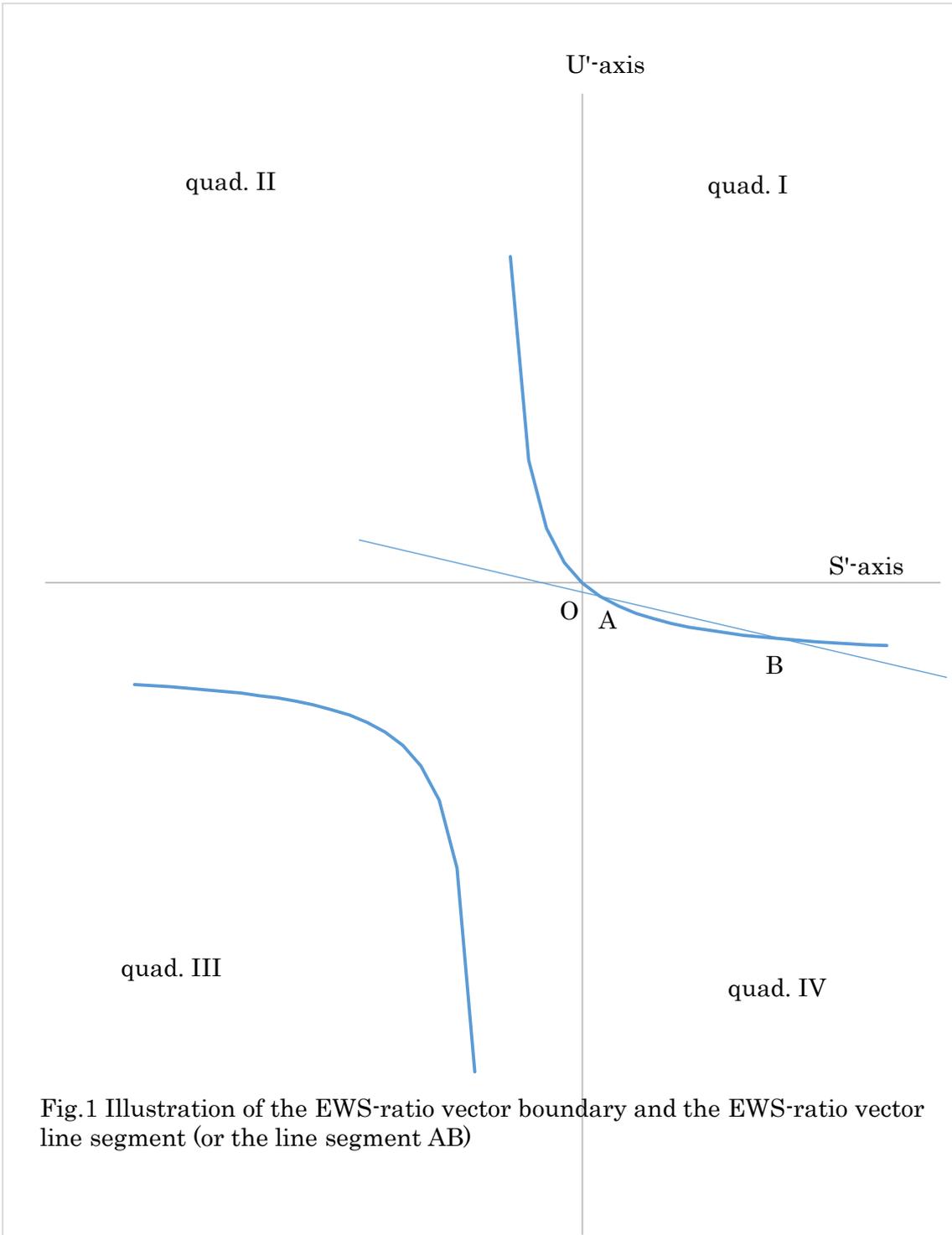

Fig.1 Illustration of the EWS-ratio vector boundary and the EWS-ratio vector line segment (or the line segment AB)

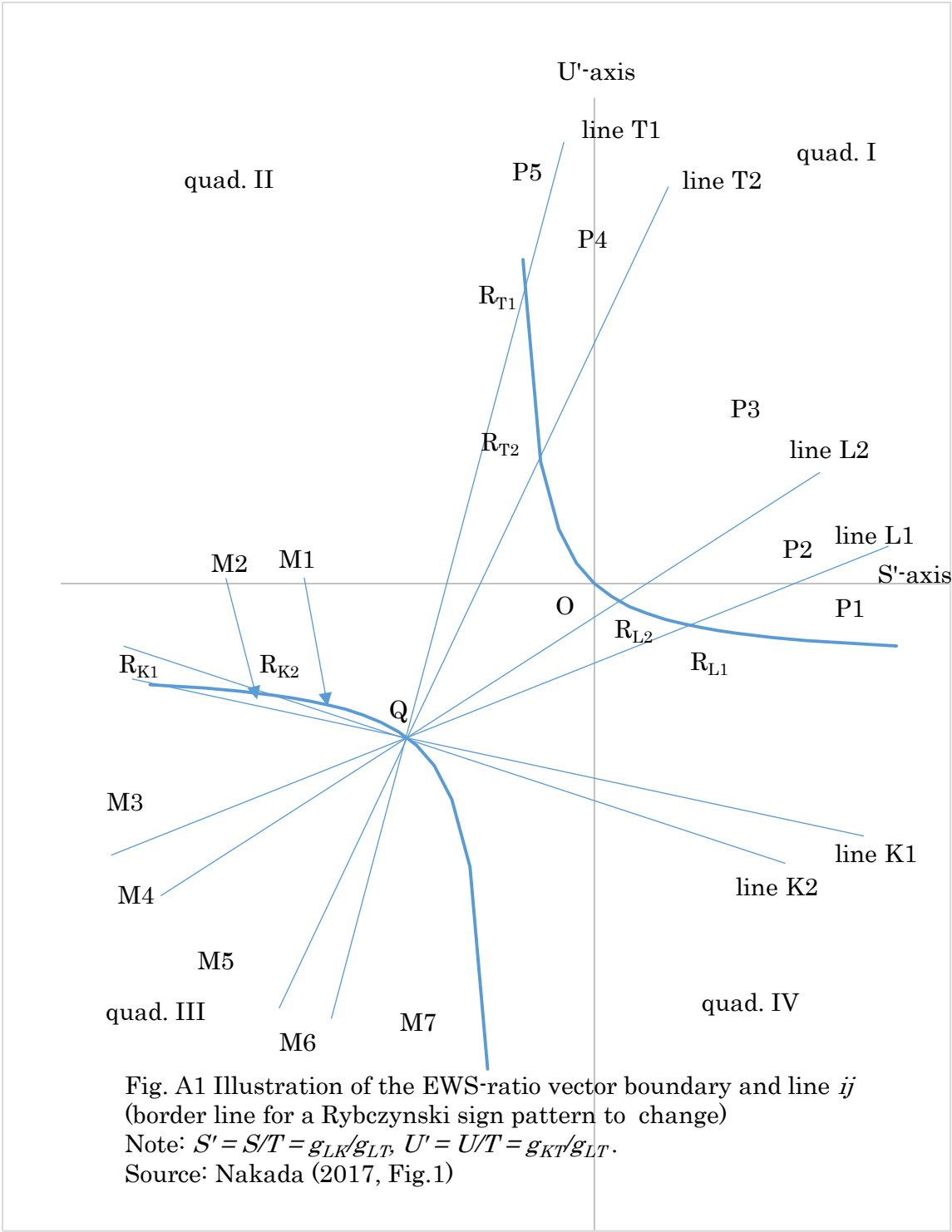

Fig. A1 Illustration of the EWS-ratio vector boundary and line $ij$
(border line for a Rybczynski sign pattern to change)
Note: $S' = S/T = g_{LK}/g_{LT}$, $U' = U/T = g_{KT}/g_{LT}$.
Source: Nakada (2017, Fig.1)

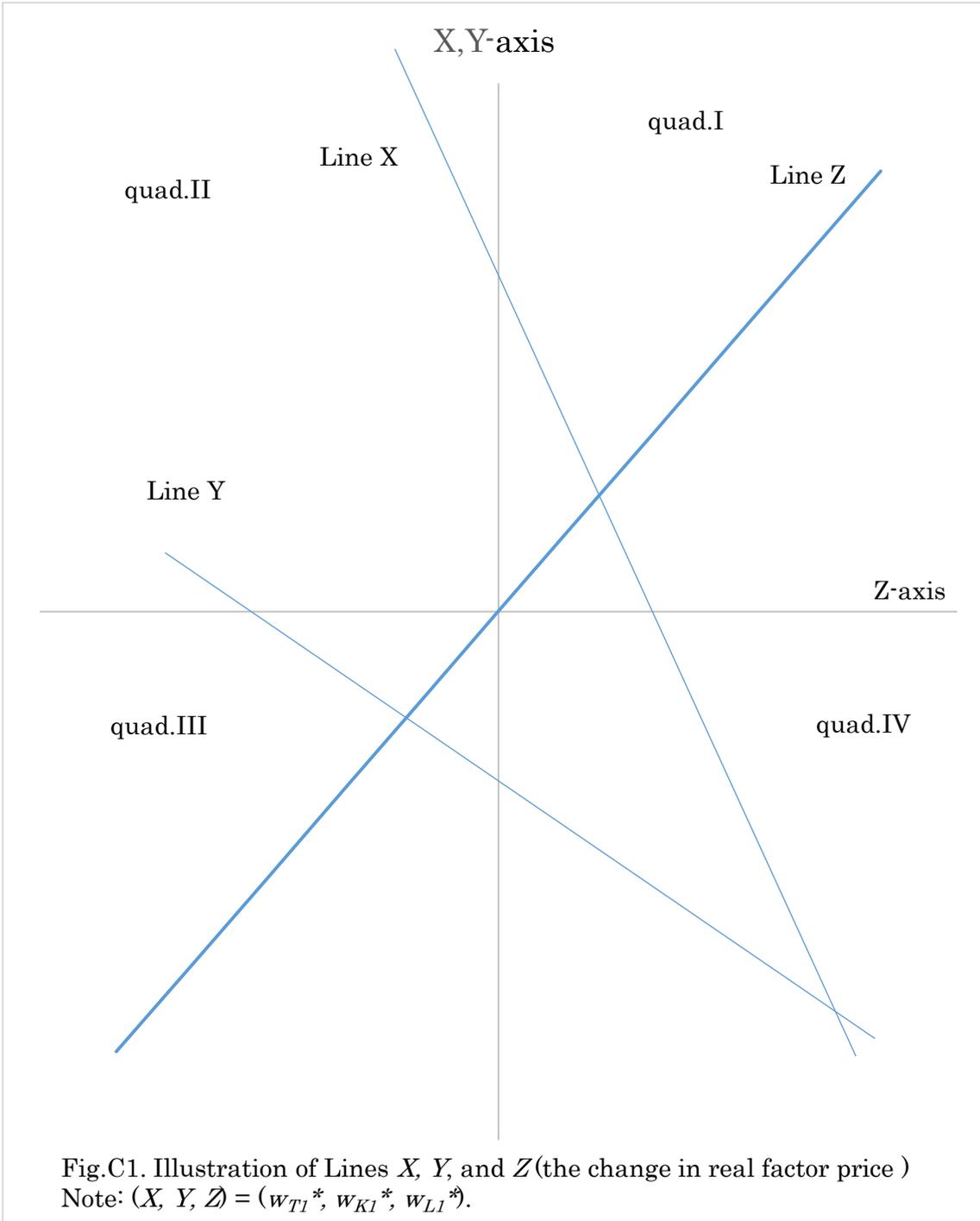

Fig.C1. Illustration of Lines *X*, *Y*, and *Z* (the change in real factor price)
Note: (*X*, *Y*, *Z*) = ($w_{T1}^*$, $w_{K1}^*$, $w_{L1}^*$).

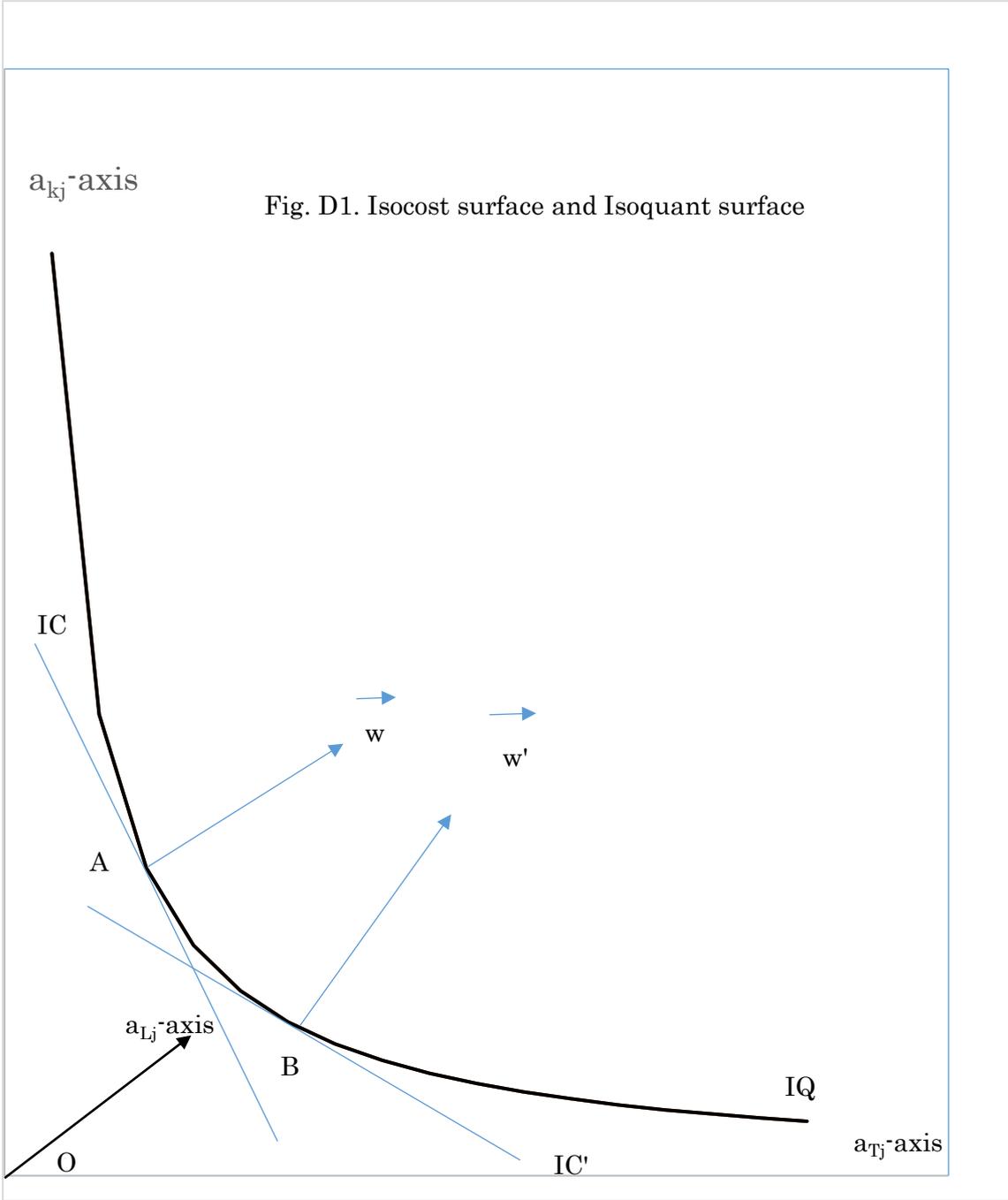

Fig. D1. Isocost surface and Isoquant surface